\documentclass[11pt]{article}
\usepackage[margin=1in]{geometry}
\usepackage[T1]{fontenc}
\usepackage{lmodern}
\usepackage[hidelinks]{hyperref}
\usepackage{microtype}
\usepackage{multicol}
\usepackage{needspace}
\usepackage{xurl}

\usepackage{fancyhdr}
\begin{document}

\begin{center}
{\LARGE\bfseries Narrative Frames: A New Approach to Analysing Metaphors in AI Ethics and Policy Discourse\par}
\vspace{1em}
Daniel Stone\par
Centre for the Future of Intelligence, University of Cambridge\linebreak
Jesus College, University of Cambridge\linebreak
Diffusion.Au\par
\href{mailto:dcs53@cam.ac.uk}{dcs53@cam.ac.uk} / \href{mailto:daniel@diffusion.au}{daniel@diffusion.au}
\end{center}

\section*{Abstract}
Metaphors fundamentally shape how we reason about complex issues like artificial intelligence, yet current approaches to metaphor analysis in political discourse suffer from inconsistent definitions and methodologies.
This paper introduces Narrative Frames, a novel categorisation system that addresses these limitations by providing a standardised framework for identifying and analysing metaphors in AI policy debates.

Building on Lakoff and Johnson's conceptual metaphor theory, we derive 49 distinct narrative frames through a two-stage process: inductively coding 685 metaphors from the MetaNet database, then cross-referencing findings with 82 critical metaphor analysis studies. This methodology grounds the typology in both empirical data and established theoretical concepts while resolving definitional ambiguities that have hindered cross-study comparison.

The Narrative Frames system offers researchers, journalists, and policymakers a shared vocabulary for analysing how metaphors shape public perception and policy priorities in AI governance. By revealing both the frames present and notably absent in discourse, this approach enables more transparent analysis of underlying assumptions and power dynamics. We discuss limitations and propose future applications, including computational scaling using large language models.

\noindent\small\textbf{Keywords:} \textit{metaphor analysis, conceptual metaphor theory, narrative frames, AI governance, political discourse, Lakoff and Johnson}

\section{Introduction}
The language we use to discuss AI is not merely descriptive; it is a powerful tool that can either illuminate or obscure the ethical stakes. Lakoff and Johnson (1980) assert metaphors are fundamental forms of reasoning, helping us think ``about abstract concepts (i.e. ideas) in terms of more concrete concepts (e.g. food)'' (Brugman et al., 2019, p.~1). They suggest as speakers select target domains (abstract concepts, 'ideas'), they highlight and hide different characteristics of target domains (concrete ideas, 'food'). Consequently, several experimental studies reveal that when strategically chosen by individuals to conceptualise issues or events, metaphors are highly persuasive, and their use has significant consequences.

For example, Boulenger (2012), Citron and Goldberg (2014) and Landau (2014) demonstrated metaphors can neurologically steer reasoning and influence behaviour. Bosman and Hagendoorn (1991), Robins and Mayer (2000) and Thibodeau and Boroditsky (2013) found metaphors effectively overcome fixed biases when used by advocates in communities with sharply divided political ideologies and affiliations. This is particularly true for complex topics like the economy, climate change, or artificial intelligence, where audiences are unfamiliar, anxious, and struggling to conceptualise implications (Landau et al., 2014; P.~Thibodeau et al., 2016).

Given its technical complexity and societal significance, public perception of AI ethics and policy is especially vulnerable to the influence of strategic framing and metaphoric reasoning as stakeholders navigate contentious debate over ethical implications. For instance, the metaphor of AI as a `race' emphasises speed, overshadowing safety and ethical concerns, potentially leading to policies favouring quick wins over long-term ethical considerations (Cave \& Ó hÉigeartaigh, 2017).

Political professionals like Westen (2008), Shenker-Osorio (2012), and Luntz (2007) have not only theorised about metaphors' impact on cognition and reasoning but also demonstrated their practical application, helping popularise this approach among the political establishment.

However, despite the crucial role metaphors play in shaping policy debates, significant concerns have emerged about the academic process of analysing them. The first section outlines multiple competing definitions of key concepts, such as `frames,' and the conflicting knowledge structures and analytical methodologies that arise from these varying definitions across different fields, revealing that these competing definitions result in misinterpretation or misallocation of potential consequences.

The second section of this paper introduces a new theoretical framework, Narrative Frames, which addresses these concerns. Grounded in Lakoff and Johnson's metaphor master list and other theoretical contributions, this framework simplifies the process of identifying metaphors and understanding the abstract concepts they represent. It also facilitates examining the relationships between alternative metaphors and evaluating their consistency with, or opposition to, a speaker's overarching goals in political communication. In doing so, it reveals the power dynamics, persuasive strategies, and intended effects when employed by influential actors.

The third section outlines the two-stage methodology employed to determine a proposed typology of narrative frames. Firstly, the 685 metaphors in the MetaNet database (Jiang et al., 2020), based on the Metaphor Master List, were coded to common themes using an inductive approach. This involved identifying recurring patterns and conceptual mappings across the metaphors and grouping them into thematically coherent categories. Secondly, a systematic cross-referencing of 82 critical metaphor analysis studies, selected based on their relevance to political discourse and use of metaphor analysis techniques, was conducted to identify common frames used in existing research. The common frames and metaphors identified through this qualitative content analysis were mapped onto themes derived from the MetaNet database, revealing 49 distinct Narrative Frames that align with Lakoff and Johnson’s theoretical foundation. This two-stage methodology ensures that the Narrative Frames categorisation system is grounded in both empirical dimensions and established theoretical concepts, enhancing its validity and applicability across various domains. In the fourth section, I discuss the limitations of the methodology and approach.

Beyond academia, Narrative Frames provide a valuable tool for journalists, policymakers, and the public to identify and analyse underlying assumptions, values, and goals promoted by different stakeholders in AI discussions. For example, journalists can use Narrative Frames to identify and deconstruct metaphors strategically employed by speakers in debates about AI regulation, advocates can more effectively map the competitive landscape between their position and their opponents', and policymakers can ensure their communication about policy implications aligns with their intentions. Together, this leads to more informed and nuanced public discourse and policy decisions that consider the complex ethical considerations surrounding AI.

In conclusion, this paper contributes to the literature on metaphors, framing, and AI ethics by proposing the Narrative Frames categorisation system. By addressing limitations of current approaches and providing a shared framework for interdisciplinary collaboration, Narrative Frames can advance our understanding of how metaphors shape public perception and policy debates related to AI. The following sections will elaborate on this approach's theoretical foundations and practical applications, situating it within the broader scholarly conversation.

\section{Major Concepts in Cognitive Linguistics}
The Narrative Frames approach builds upon existing cognitive linguistic and discourse theory that explores the importance and influence of metaphor and framing techniques on cognition. After reviewing the existing literature, relevant concepts can be categorised into three levels:

\begin{itemize}
\item A Conceptual level (Broader, abstract ideas and cognitive processes),
\item A Systematic level (Structured, organised aspects of cognition),
\item A Technical level (Specific ways knowledge is represented and organised).
\end{itemize}

\subsection{Conceptual Level}
The \textbf{conceptual level} encompasses broad, abstract ideas and cognitive processes that shape our understanding. Key concepts include Johnson's ``Image Schemas'' or ``cogs,'' which are pre-linguistic embodied structures described as ``a recurring, dynamic pattern of [perception] that give coherence and structure to our experience'' operating ``above that of concrete rich images'' (1987, p.~xiv). These ideas about associative meaning-making relate to Kant's transcendental schema (Hanna, 2017; Kant, 2003).

Later, Johnson and Lakoff (1980) introduced `experiential gestalts,' which are similar to image schemas but more complex, multidimensional images arising from bodily interactions, linguistic experience, and historical or cultural contexts. Image schemas initiate them but contribute other components like participants, parts, and stages clustered ``to form a gestalt---a whole that we human beings find more basic than the parts'' (Lakoff \& Johnson, 1980, p.~70).

The concept of gestalts, for instance, serves the medical hypothesis that ``more complex conceptual structures are the result of neural networks that simultaneously activate multiple bound primitive structures located in various areas of the brain'' (Feldman, 2006; Feldman \& Narayanan, 2004; Lakoff, 2008; as cited in Stickles et al., 2016, p.~56). These neural structures ``are activated as part of the overall spreading of neural activation'', directly influencing thought processes.

Quinn and Holland (1987) highlight that experiential gestalts, the most interpretive and abstract layer of Lakoff and Johnson's work, are particularly relevant in studying the contextual and associative role of culture in political reasoning. This is exemplified by Lakoff and Johnson's suggestion that the experiential gestalt of ``argument'' is often understood through the lens of ``war.''

The idea that cognition begins at an unstructured and associative level is reflected in similar concepts such as Nikoli's (2009) Ideasthesia, where the activation of certain concepts triggers sensory experiences similar to actual perceptions; Piaget's (1952) theory of schema, which involves hierarchically categorised units of understanding existing in complex relationships; and Fauconnier and Turner's (2003) Conceptual Blending, where elements and relations between different scenarios are subconsciously ``blended'' as a ubiquitous process in everyday thought and language.

These concepts highlight the abstract and associative stage, which begins cognitive processes, and the role of embodied experience in shaping conceptual understanding.

\subsection{Systematic Level}
The \textbf{systematic level} focuses on the structured, organised aspects of cognition, where the contested definition of `frames' comes into view. Originating in Minsky's (1975) work on knowledge representation, `frames' have become a fundamental, albeit sometimes ambiguous, notion across various fields, including cognitive semantics, computational linguistics, artificial intelligence, cognitive psychology, media studies, and discourse studies (Ziem, 2021 for an overview). Despite differing definitions and views on their function and influence, two common conceptualisations of frames are popular in the study of political language.

\Needspace{18\baselineskip}
\subsubsection{Semantic Frames}
\nopagebreak[4]
Fillmore's (1975, 1976, 2001) Semantic Frames are schematic representations of situations or events, including participants, props, and roles, that provide the background and cognitive structure necessary for understanding word meanings. A word cannot be understood without access to all essential related knowledge; for example, `buy' is interconnected with `sell', `cost', and `spend'. When a word is used, it evokes a semantic frame of encyclopaedic meaning from a particular, culturally-defined perspective. Fillmore (1975, p.~128) illustrates this with the example of ``bachelor,'' asking, ``Why is it confusing to ask if the pope is a bachelor?'' Holland and Quinn (1987, p.~23) explain that ``the word bachelor 'frames' a simplified world in which prototypical events unfold,'' and ``The Pope, under a vow of celibacy, is not relevant character in the cultural world of marriage.'' Also known as ``scripts'' or ``schemata,'' semantic frames have significantly contributed to natural language understanding research (Schank \& Abelson, 1977) and the development of Large Language Models.

Fillmore's work forms the basis for several narrative analytic approaches. For instance, Chambers and Jurafsky's ``narrative event chains'' (2008, p.~789) have become a popular narrative discourse tool by introducing temporal and causal dimensions to semantic frames, defining them as ``a partially ordered set of events related by a common protagonist.'' While semantic frames provide background knowledge for understanding individual words and events, narrative event chains capture the higher-level structure and coherence by linking frames together based on a common protagonist and a sequential, causal relationship.

\subsubsection{Conceptual Metaphor Frames}
However, for analysing discourse's ethical and political dimensions, George Lakoff and Mark Johnson’s Conceptual Metaphor Frames, introduced in \textit{Metaphors We Live By} (1980), offer a more directly relevant approach. These metaphorical frames allow us to formalise their idea of gestalts by mapping them conceptually between a source domain (the more tangible, experiential concept) and a target domain (the abstract concept used to aid understanding). By highlighting specific aspects of the target domain and obscuring others, these frames shape our normative views and influence how we perceive and judge political issues.

Thus, while Semantic Frames provide the structural background for understanding how specific language shapes perception, Conceptual Metaphor Frames are pivotal in shaping our ethical and political interpretations. The latter's perceptual structuring makes it a foundational approach in analysing how language reflects and shapes our understanding of policy debates.

A relevant idea in conceptual metaphor theory, `entailments' structure cognition and operationalise abstract gestalts. They are not random or isolated words but consistent mappings that translate inferential logic from the source to the target domain. For instance, when the two earlier gestalts bond in the `ARGUMENT is WAR' metaphor, it uses entailments evoking battle strategies, attacks (`he shot down my point'), defences (`I defended my argument'), and outcomes of winning (`I was victorious') or losing (`I was defeated'). The cumulative use of entailments over a text transforms abstract concepts into comprehensible images and familiar stories, building a coherent conceptual picture that shapes our understanding of the target concept and distinguishes metaphorical mappings from simple literary devices.

Lakoff and Johnson's `entailments' operate at a more abstract level than Fillmore's ``scripts'' and Chambers and Jurafsky's ``narrative event chains''. While Lakoff and Johnson do not extensively discuss culture, Kövecses points out that it influences entailments. However, Lakoff and Johnson argue that the cognitive processes underlying entailments are universal, suggesting that deep cognitive structures of metaphorical mappings are shared across cultures, even if specific mappings are shaped by cultural context.

Lakoff and Johnson show that consistent metaphorical mappings shape our understanding by highlighting and obscuring aspects of the target domain, profoundly influencing how we perceive and judge political issues. Consequently, conceptual metaphor theory has been embraced across disciplines like cognitive linguistics, philosophy, psychology, computer science (notably machine learning), and political science (Colburn \& Shute, 2008; Gibbs, 1994, 2011; Kovecses, 2010). However, each field interprets and applies these frameworks through its own lens, often leading to divergent interpretations that can challenge cross-disciplinary integration and dialogue.

\Needspace{10\baselineskip}
\subsection{Technical Level}
The \textbf{technical level}, the most concrete of the three levels, focuses on the specific mechanisms and structures used to represent and organise knowledge in the mind. This is the `pointy end', where researchers have developed methodologies, typologies, and diagnostic tools to investigate how information is stored, retrieved, and manipulated within cognitive systems, as well as the consequences and implications of its use. My focus narrows to aspects of Lakoff and Johnson's conceptual metaphor theory most relevant to understanding the construction and persuasiveness of political discourse while acknowledging related theories (such as Fillmore’s) where appropriate.

Lakoff and Johnson's (1980) master list of conceptual metaphors has become an influential reference point for studying framing and metaphor. The list provides a systematic typology of recurring mappings between source and target domains, such as ``CHANGE is MOTION'' and ``HOPE is LIGHT''. By organising conceptual metaphors based on their underlying structure, the master list offers a rigorous framework for understanding how metaphors operate across abstract concepts.

However, the master list has faced scrutiny. Vervaeke and Kennedy (1996) argue that it oversimplifies metaphor and fails to capture its complexity and diversity. Kövecses (2010) and Idström and Piirainen (2013) challenge its cultural bias towards Western, English-speaking cultures. Gibbs (2011) and Steen (2011) question its intuitive nature and call for empirical validation through corpus linguistics and experimental studies. Semino (2008) and Musloff (2004) suggest it overlooks domain-dependent metaphors, particularly in political discourse. Goatly (1997) argues it overemphasises universality while downplaying cultural, historical, and ideological factors, and Charteris-Black (2011) calls for more attention to the social-political dimensions and power relations in metaphor use.

Despite this, the master list's structured approach has made it an invaluable tool across different fields. It provides a shared vocabulary, well-defined rules, and a systematic framework that facilitates its application, adaptation, and consistent use by researchers (including all of the critical authors above). Despite its age, it remains a foundational resource for many empirical and theoretical approaches to uncovering the underlying conceptual structures that guide our thinking about complex political issues.

These approaches generally divide into two distinct paths of practice: the turn towards predictable empiricism and the path towards critical theory.

\subsubsection{The Turn Towards Predictable Empiricism}
Lakoff and Johnson's attempt to establish syntactic mappings has become attractive to empirically-minded researchers in linguistics, social science, and computer science.

The metaphor master list's typology of metaphorical mappings has inspired a substantial body of schematic literature and is the foundation for several metaphor databases, including MetaBank (Martin, 1994), ATT-Meta (Lee \& Barnden, 2001a, 2001b), FrameNet (Baker et al., 1998; Fillmore et al., 2003; Litkowski, 2010), and MetaNet (Stickles et al., 2016).

The most notable are FrameNet, which focuses on the syntactic properties of semantic frames, and MetaNet, which specifically addresses metaphorical frames and the mappings between them. These resources support various forms of discourse and narrative analysis, such as examining metaphorical framing of news discourse (Stickles et al., 2016), by providing a structured way to explore language's conceptual underpinnings.

Similarly, efforts have been made to identify the relationship between formal metaphors and neural activations.

\Needspace{8\baselineskip}
\subsubsection{Critical Approaches}
Notable contributions include Critical Metaphor Analysis (CMA), developed by scholars like Jonathan Charteris-Black (2004) and Andreas Musolff (2011), which critically examines the ideological and power dynamics behind metaphor use. Musolff (2006) proposed the Metaphor Scenario Analysis approach, which delves into the narrative stories that metaphors draw upon and contribute to, offering insights into how metaphors shape and are shaped by these narratives. Meanwhile, the Discourse Dynamics Framework, developed by Cameron et al. (2009), emphasises interactional dynamics and how metaphors evolve dynamically in real-time conversation rather than their socio-political implications.

These methods involve three stages: identification (which metaphors are present, using techniques from the technical level), interpretation (what usage patterns and constructions they support or undermine, using concepts from the systematic level), and explanation (who uses them, how they interact with context, and the ideological reasons for their use, typically drawing from the conceptual level). While these critical approaches offer valuable insights into the role of metaphors in shaping public discourse, their application in practice raises some methodological concerns.

Researchers often develop a study-specific typology during the identification and interpretation stages, sometimes loosely based on the metaphor master list. However, identifying themes within the text remains largely subjective and reactive to the specific text under analysis (Semino et al., 2018). Despite implying a technical and empirical base, this approach often simplifies master list concepts without a transparent methodology, causing inconsistencies and a lack of consensus among researchers (Cacciatore et al., 2016; Leeper \& Slothuus, 2020). This is exacerbated by the lack of a consistent definition of `frames' (Matthes, 2009; Scheufele \& Iyengar, 2014), which can lead to attributing effects to frames that diverge from their intended theoretical functions (Herrnson et al., 2007; Hertog \& Mcleod, 2001). Referencing experimental studies from other fields can enhance the perceived legitimacy of claims about discourse impacts, but these claims must be based on a clear and precise understanding of the underlying concepts (Borah, 2011; Entman et al., 2008). Otherwise, assertions about cognitive processes and their socio-political implications risk being misleading.

To bolster causal claims and enhance interdisciplinary dialogue, especially in public policy and AI, researchers using critical metaphor analysis methods must integrate their work with insights from linguistics, psychology, neuroscience, and computer science. This multidisciplinary approach is necessary to fully understand how metaphors influence cognition, communication, and behaviour in political contexts (Gibbs Jr., 2017; Thibodeau et al., 2017). Creating a shared understanding of key terms across disciplines will foster more rigorous, empirically grounded research, enhance visibility, facilitate collaboration, and pave the way for advanced interpretive computational models (Shutova et al., 2016; Veale et al., 2016).

\section{Narrative Frames}
This paper introduces `Narrative Frames', a novel categorisation model that bridges the divide between metaphorical framing in policy discourse practice and existing conceptual linguistic frameworks. It addresses some limitations in critical metaphor analysis by deriving a common set of narratives directly from the original metaphor master list. This reduces subjectivity and inconsistency in the identification and interpretation of metaphors while providing a more accessible and structured approach for researchers in policy and political rhetoric. Moreover, Narrative Frames conceptually reinforces the connection between discourse and cognitive processes, better contextualising the strength and limits of any identified effects and facilitating enhanced comparative analysis between studies.

\subsection{Conceptual Foundation}
Narrative Frames\footnote{This has no relationship to Barkhuizen’s (2008) `narrative frames’, which are qualitative diagnostic tools typically used in education.} are conceptually defined according to Entman's (1993, p.~52) interpretation of framing, which highlights specific problems, causal relationships, moral evaluations, and solutions while simultaneously obscuring others, directing the audience's understanding in a particular way. The foundation of Narrative Frames lies in Johnson and Lakoff's work on Conceptual Metaphors (often called Metaphoric Frames). However, Narrative Frames diverge from this foundation in two key aspects:

Firstly, Narrative Frames broaden the scope of reference beyond strict focus on linguistic metaphors to include analogies, similes, and other forms of figurative language that use non-literal language to communicate abstract concepts. This broader scope is identified using a combination of Charteris-Black's (2004) and Pragglejaz’s (2007) methodologies, as proposed by Imani (2022), for identifying metaphoric language and developing a corpus for study.

Secondly, Narrative Frames directly evolve the metaphor master list into a typology of common narratives derived from the list itself. The metaphor master list, with its narrow mathematical mapping structure, tends to deconstruct conceptual metaphors into their components without placing them in the broader narratives they support or propagate. Narrative Frames address this by developing a typology echoing Musloff's (2006) ``metaphorical scenarios'' and Werth's (1994) ``mega-metaphors'', which are extended metaphorical mappings that unfold over the course of a text or discourse.

By broadening the field of view to all non-literal language and acknowledging the compounding and associative nature of meaning-making across a text, Narrative Frames bring back the associative dimension to cognition outlined in gestalts and operationalised through entailments. This approach reinforces the connection between discourse and cognitive processes, better contextualising the strength and limits of any identified effects and facilitating enhanced comparative analysis between studies.

The term ``narrative'' represents these frames because it captures the holistic and associative nature of meaning-making that occurs at the gestalt level in Lakoff's theory. Narratives are powerful tools for organising and conveying complex information, as they provide a coherent structure that integrates various elements, such as events, characters, and themes, into a meaningful whole (Bruner, 1991; Fisher, 1984). This narrative structure closely resembles the way gestalts operate in Lakoff's framework, where multiple components (i.e. participants, parts, and stages) are integrated into a coherent whole (Lakoff \& Johnson, 1999). Moreover, policy discourse commonly uses narratives to frame issues, shape public opinion, and influence decision-making (Jones \& McBeth, 2010; Shanahan, Jones, \& McBeth, 2011). Aligning frames with narratives helps better understand and analyse language's role in shaping policy discourse and its effects on public perception and behaviour.

Aligning with our approach, Lakoff and Johnson themselves took a simpler, more context-sensitive approach to metaphor when communicating their work to a broader audience. In \textit{Don't Think of an Elephant} (2004), they discuss concepts similar to WAR and FAMILY without strict definitions, using narrative techniques to make their ideas more relatable and engaging for less specialised readers.

Narrative Frames acknowledges that the repeated, contextually relevant deployment of Lakoff and Johnson's Conceptual Metaphors elevates them from being one step in a linguistic reasoning process. In doing so, it helps us maintain a clear link to their historical work and the fields that have emerged from it while also answering how these metaphors seek to address the questions, ``What am I to believe? And what am I to do?''

\section{Methodology: Identifying Narrative Frames}
The proposed list of narrative frames was developed through two complementary processes.

\subsection{Identifying and Classifying Target Domains}
The 685 metaphors currently listed in the MetaNet database, which has continued the work started by Lakoff and Johnson's original metaphor master list, were divided into their source domain (X) and target domain (Y) components using the `is' or `are' keywords as a separator. For example, the metaphor `ARGUMENT is WAR' was divided into the source domain `ARGUMENT' and the target domain `WAR'. This process allowed for the isolation and examination of the target domains, which represent the conceptual domain we aim to understand through the metaphor.

An inductive coding approach was conducted manually to group similar target domains into broader narrative frames with similar narrative strategies, imagery or themes. The target domains were examined to identify common themes, concepts, or meanings that united them. These narrative frames were not predetermined but rather emerged from the analysis of the data. For example, the target domains `WAR', `ATTACKING', `BATTLEFIELD', `INVADING A TERRITORY' and `DEFENDING A TERRITORY' were grouped into a broader \texttt{WAR} narrative frame based on their shared conceptual meanings.

Some of the MetaNet metaphors presented multiple or unclear target domains during the coding process. For example, `FUTURE OF A NATION is THE PATH OF A BOAT' could be coded into either the proposed \texttt{JOURNEY} or \texttt{MACHINE/TRANSIT} narrative frames. In such cases, the most salient aspect of the metaphor was prioritised, with the `path of a boat' being coded as \texttt{JOURNEY} based on the emphasis on movement and direction and active imagery of holidays or migration. Similarly, ``GOVERNMENT'S HARMFUL TREATMENT OF RIGHTS is DAMAGING A PHYSICAL STRUCTURE'' could be interpreted as \texttt{BUILDING} or \texttt{WAR}, depending on the context of use. The subject of the activity, \texttt{BUILDING}, was given precedence in this decision. These examples illustrate the nuanced approach to resolving ambiguous cases, guided by salience and contextual relevance.

The process of defining and refining the narrative frames was iterative. As more target domains were coded, the existing frames were continuously compared and modified to better capture the conceptual relationships between domains and to form clear definitions. This involved expanding or narrowing the scope of frames, creating new frames when necessary, and resolving any challenges in defining the boundaries between frames.

The iterative process produced 49 narrative frames, each representing a unique thematic connection within the MetaNet metaphors. This approach refines the master metaphor list, articulates broader themes in conceptual metaphor research, and bridges the gap between traditional metaphor analysis and discourse analysis methods. By aligning with common terminologies used by discourse analysts, the Narrative Frames typology enhances its validity and applicability.

\subsection{Integrating and Refining Narrative Frames}
To validate and refine the proposed narrative frame typology, a systematic cross-referencing was done with existing critical metaphor studies to identify commonly used labels relating to political and policy communication. This establishes consistency between the inductively derived frames and the findings of previous studies, identifies potentially overlooked frames, and situates the typology within broader metaphorical framing research.

Research literature from 1980\footnote{The year Lakoff and Johnson published the master list.} to 2023 was selected via JSTOR and Google Scholar, focusing on peer-reviewed English articles in three main categories:

\begin{enumerate}
\item Works that cite Lakoff and Johnson's seminal work \textit{Metaphors We Live By} (1980), include metaphors from the master list and contain the words `frame,' `policy,' `rhetoric,' or `discourse' in their title or abstract.
\item Studies that cite the original papers proposing the three most common critical analytical techniques for metaphor analysis: ``Discourse Dynamics Framework'' (Cameron et al., 2009), ``Metaphor Scenario Analysis'' (Musolff, 2006), and ``Critical Metaphor Analysis'' (Charteris-Black, 2004).
\item Works that attempt to develop a list of typologies discovered as part of the broader literature review for this paper.
\end{enumerate}

To focus on metaphorical framing in real-world political communication, papers were screened to include only studies that apply the mentioned techniques or build on Lakoff and Johnson's work to analyse political discourse from primary data (i.e. speeches, policy documents, media coverage), including meta-reviews synthesising such studies. Primarily theoretical or methodological articles were excluded.

A qualitative content analysis using a directed approach (Hsieh \& Shannon, 2005) was then conducted on the 82 qualifying papers\footnote{See Appendix for citations.} to identify metaphors in the selected texts and allocate them to the categories in the draft typology. The directed approach was chosen as it allows for validating and extending an existing theoretical framework or typology.

The analysis began with a close reading of each qualifying paper's title, abstract, and methodology section, as these typically contain the most explicit and concise information about the key terms and definitions. During this reading, specific instances of metaphorical language were identified and recorded.

Significant commonalities began to emerge. For example, Charteris-Black (2004) identified metaphors he labelled \texttt{CONFLICT, JOURNEY, HEALTH, BUILDING, FIRE, PHYSICAL ENVIRONMENT, RELIGION}, and \texttt{BODY PARTS}. Cammaerts (2012) identified \texttt{SPORTS, GAMES, WAR}, and \texttt{TRANSPORT}. Rotimi Taiwo (2010) found \texttt{BUILDING, BATTLE} and \texttt{JOURNEY}. Vestermark (2007) saw a \texttt{COMMUNITY, PERSON} and \texttt{HUMAN BEING}. Mouna Hamrita (2016) found \texttt{ALIENATION, EVIL, DEMON} and \texttt{WAR}. Ezeifeka (2013) identified \texttt{SACRIFICE, REWARDS IN HEAVEN, WAR/CONFLICT, VILLAINS, DISEASE/ILLNESS, A COG IN THE WHEEL}, and \texttt{DEAD ISSUE}.

For example, despite the use of `conflict', `war', and `battle', the dynamics each author associates with the label are often very similar and, in many cases, use the same entailments. Charteris-Black (2011, p.~169) writes of his `conflict' metaphor, ``[they] are used to frame government policies as a military campaign''. Cammaerts (2012, p.~10) describes their `war' metaphor as ``political actors being on the defensive, digging in \dots{} denoting an unwillingness to compromise while at the same time waging a full-scale political war \dots{} requiring a united front against the enemy.'' Taiwo (2013, p.~201) describes `battle' as a ``conceptualisation of [social] problems as something to war against \dots{} in a battle to establish freedom from dominant forces''. Hamrita (2016, p.~100) identifies `war' when a politician situates opponents as ``an enemy'' in ``clash and conflicts'' where the public is encouraged to ``fight them as they represent a symbolic destructive threat to the Tunisian identity''. Ezeifeka (2013, p.~184) wrote that `war/conflict' frames ``appeal to the emotions\dots{}about devastating consequences'', ``which will take their toll on innocent victims'', who must be ``protected and defended''.

Identified metaphors were allocated to corresponding categories in the draft typology based on their conceptual similarity. For example, the above examples were evidence of a `WAR' narrative frame, which appears in the master list as ``ARGUMENTS are WAR''. As conceptual nuance began to emerge within common frames, nested, secondary narrative frames were created (such as \texttt{JOURNEY/QUEST} and \texttt{JOURNEY/RACE}) to reflect the distinct narrative dynamics that emerged.

Few existing works have tried to develop the master list into a typology similar to narrative frames. Kövecses (2010) perhaps making the most notable contribution, clustered the metaphors into 13 categories: \texttt{ANIMALS}, \texttt{BUILDINGS AND CONSTRUCTION}, \texttt{COOKING AND FOOD}, \texttt{FORCES}, \texttt{GAMES AND SPORTS}, \texttt{HEALTH AND ILLNESS}, \texttt{HEAT AND COLD}, \texttt{LIGHT AND DARKNESS}, \texttt{MACHINES AND TOOLS}, \texttt{MONEY AND ECONOMIC TRANSACTIONS}, \texttt{MOVEMENT AND DIRECTION}, \texttt{PLANTS}, and \texttt{HUMAN BODY}. Yet there are some unusual exceptions---\texttt{JOURNEY}, \texttt{WAR} and \texttt{SUBSTANCE}, explicitly stated in the master list and commonly cited in the literature, are notably absent.

The process of identifying metaphors, allocating them to categories, and refining the typology was iterative and recursive. As new metaphors and categories emerged, the typology was renegotiated and restructured to capture better the range of metaphors present in the selected texts while also striving to maintain a concise and manageable set of categories to facilitate the typology's use as a practical tool for researchers. This back-and-forth process continued until a comprehensive and coherent typology was developed that accounted for all identified metaphors.

The final typology, resulting from this iterative analysis, not only validates the initial categories derived from the metaphor master list but also extends and refines them based on the metaphors used in real-world discourse.

\section{Proposed Schema of Narrative Frames}
The systematic process described in the previous section identified 49 narrative frames, organised into a hierarchical framework that delineates primary conceptual domains (e.g., \texttt{JOURNEY}, \texttt{MACHINE}) and nested categories reflecting specific instantiations with distinct narrative implications (e.g., \texttt{JOURNEY/RACE}, \texttt{JOURNEY/QUEST}). This typology captures the diverse ways metaphorical framing operates in political discourse to shape public opinion and policy debates.

The narrative frames in this typology serve as the fundamental foundations of metaphoric narratives, each with its unique characteristics and implications. Some frames, such as \texttt{JOURNEY}, \texttt{WAR}, and \texttt{GAME}, offer rich narrative landscapes that allow for complex storytelling and analogies, evoking specific actors, dynamics, goals, and challenges. Others, like \texttt{BODY}, \texttt{SUBSTANCE}, and \texttt{OBJECT}, focus on single aspects or components of a concept, representing a more elemental form of storytelling with limited scope for narrative elaboration. Frames like \texttt{MACHINE}, \texttt{HEALTHCARE/DISEASE}, and \texttt{BUILDING/CONSTRUCTION} facilitate discussions about how specific actions can have specific effects, such as `pulling a policy lever' or `firming up public support.'

It is important to note that while these narrative frames are the building blocks of metaphoric narratives, they are not exhaustive and can be integrated with each other to create new meanings. Narrative frames can be paired to create ``blended spaces'' (Fauconnier \& Turner, 2003), enhancing meaning and emphasising specific characteristics. For example, blending the \texttt{WAR} and \texttt{RACE} frames can form an ``Arms Race'' narrative, combining elements from different frames to create a cohesive story. However, in these blends, the fundamental characteristics of the original frames persist, maintaining their distinct conceptual logic and narrative implications.

The following Narrative Frame Schema, listed in no particular order, provides a preview of the full typology:

\Needspace{30\baselineskip}
\subsection{Narrative Frame Schema}
\begin{multicols}{3}
\raggedcolumns
\footnotesize
\setlength{\columnsep}{1em}
\begin{enumerate}
\setlength{\itemsep}{0pt}
\setlength{\parskip}{0pt}
\item \texttt{TIME}
\item \texttt{ANIMAL}
\begin{enumerate}
\setlength{\itemsep}{0pt}
\item \texttt{ANIMAL/BEAST}
\end{enumerate}
\item \texttt{WAR}
\item \texttt{GAME}
\item \texttt{JOURNEY}
\begin{enumerate}
\setlength{\itemsep}{0pt}
\item \texttt{JOURNEY/RACE}
\item \texttt{JOURNEY/QUEST}
\item \texttt{JOURNEY/SPATIAL}
\end{enumerate}
\item \texttt{MACHINE}
\begin{enumerate}
\setlength{\itemsep}{0pt}
\item \texttt{MACHINE/CAR}
\item \texttt{MACHINE/TRANSIT}
\end{enumerate}
\item \texttt{HEALTHCARE}
\begin{enumerate}
\setlength{\itemsep}{0pt}
\item \texttt{HEALTHCARE/DISEASE}
\item \texttt{HEALTHCARE/VIRUS}
\end{enumerate}
\item \texttt{BODY}
\begin{enumerate}
\setlength{\itemsep}{0pt}
\item \texttt{BODY/EMBODIMENT}
\item \texttt{BODY/SENSORY}
\end{enumerate}
\item \texttt{NATURAL WORLD}
\begin{enumerate}
\setlength{\itemsep}{0pt}
\item \texttt{NATURAL WORLD/PLANT}
\item \texttt{NATURAL WORLD/ECOSYSTEM}
\item \texttt{NATURAL WORLD/WEATHER}
\end{enumerate}
\item \texttt{SUBSTANCE}
\item \texttt{OBJECT}
\begin{enumerate}
\setlength{\itemsep}{0pt}
\item \texttt{OBJECT/CONTAINER}
\end{enumerate}
\item \texttt{BUILDING}
\begin{enumerate}
\setlength{\itemsep}{0pt}
\item \texttt{BUILDING/CONSTRUCTION}
\end{enumerate}
\item \texttt{TRANSACTION}
\item \texttt{INAUTHENTIC}
\item \texttt{FOOD AND COOKING}
\item \texttt{MYTHICAL}
\begin{enumerate}
\setlength{\itemsep}{0pt}
\item \texttt{MYTHICAL/JUDEOCHRISTIAN}
\item \texttt{MYTHICAL/CLASSICAL}
\item \texttt{MYTHICAL/SPIRITUAL}
\end{enumerate}
\item \texttt{VIEW}
\begin{enumerate}
\setlength{\itemsep}{0pt}
\item \texttt{VIEW/WINDOW}
\item \texttt{VIEW/CAMERA}
\end{enumerate}
\item \texttt{PERFORMANCE}
\begin{enumerate}
\setlength{\itemsep}{0pt}
\item \texttt{PERFORMANCE/THEATRE}
\item \texttt{PERFORMANCE/VISUAL ARTS}
\item \texttt{PERFORMANCE/WRITING}
\item \texttt{PERFORMANCE/MUSIC}
\end{enumerate}
\item \texttt{POLICING}
\item \texttt{LEGAL ORDER}
\item \texttt{POWER AND HIERARCHY}
\begin{enumerate}
\setlength{\itemsep}{0pt}
\item \texttt{POWER AND HIERARCHY/WORKPLACE}
\item \texttt{POWER AND HIERARCHY/SLAVERY}
\end{enumerate}
\item \texttt{RELATIONSHIPS}
\begin{enumerate}
\setlength{\itemsep}{0pt}
\item \texttt{RELATIONSHIPS/FAMILY}
\end{enumerate}
\end{enumerate}
\end{multicols}

The following sections will explore two case study narrative frames, demonstrating the typology's value as a comprehensive tool for understanding and critiquing metaphorical language's power in political discourse.

\subsection{A Closer Look: War}
The \texttt{WAR} narrative frame is highly oppositional and competitive, presenting challenges, conflicts, or efforts as military battles. It places the reader in relation to an ``opponent'', presenting an adversarial perspective that focuses on the role of strategy choices in achieving a goal. This is high stakes, zero-sum and defined by victory or defeat.

It often emphasises a core set of `good' values or principles we must rally around and defend to mobilise united, collective action with a burning sense of urgency and focus. It is characterised by entailments drawn from military language (i.e. battle, front lines, fight, invasion, adversary) and military symbols, imagery or references. This effect can be strengthened by using language that, while not overtly militaristic, enhances the frame in context, such as phrases like `stand together' and `take action'.

Its usage may have several normative ethical implications for AI policy. Firstly, it may lead to polarisation and adversarial dynamics (hindering constructive dialogue, collaboration, or compromise). Secondly, it may oversimplify policy challenges' complex and multifaceted nature into `good vs bad' discussions. Thirdly, it may lead to the justification of extreme measures and overt militarisation of the technology (i.e. ignoring ethical principles such as transparency and accountability or directing funding towards R\&D in military contexts rather than civilian ones).

However, in certain contexts, the \texttt{WAR} frame may serve to underscore the gravity of the situation and the need for decisive action. By framing AI policy challenges as existential threats, may galvanise public support and political will to address pressing issues. The frame's emphasis on unity and collective action could also foster a sense of shared responsibility and purpose in tackling complex problems.

\Needspace{10\baselineskip}
\Needspace{10\baselineskip}
Examples include:
\begin{itemize}
\item ``The \textbf{war} on unethical AI'',
\item ``On \textbf{the front lines} of AI Ethics'',
\item ``We must \textbf{marshal} our efforts and \textbf{combat} climate change'',
\item ``\textbf{Battleground} of data privacy'',
\item ``\textbf{Deploying} AI Safety'',
\item ``Unregulated AI deployment is a \textbf{minefield}'',
\item ``We’re being \textbf{blitzed} by open-source models''.
\end{itemize}

As one of the most common frames, \texttt{WAR} frequently blends with other frames that suggest momentum (e.g., \texttt{JOURNEY} or \texttt{RACE}) or cause and effect (e.g., \texttt{MACHINE}, \texttt{BUILDING}). It shares many competitive and oppositional mechanics with the friendlier and more cooperative \texttt{GAME}.

\subsection{A Closer Look: Journey}
The Journey narrative frame is a progressive and exploratory framework that situates challenges or pursuits within the context of a metaphorical expedition towards a desired outcome. It places the reader as an agent of movement in a dynamic spatial environment, emphasising strategic navigation and overcoming obstacles, whether natural, manufactured, or adversarial.

It contains three sub-variants (detailed as distinct concepts in the supplementary schema) that represent the unique characteristics of this frame. Firstly, \texttt{JOURNEY/RACE} (which portrays progress or goal pursuit as a competitive race, emphasising speed, direct competition, rules of engagement, and rewards for success), \texttt{JOURNEY/QUEST} (which depicts progress or goal pursuit as a quest, emphasising a deep sense of purpose, overcoming significant hardships, and the endurance required to achieve a meaningful goal), and \texttt{JOURNEY/SPATIAL} (which focuses on the reader's spatial relationship to other entities or locations without the implication of a goal-directed journey).

Looking at the \texttt{JOURNEY/RACE} narrative frame more closely, it is a dynamic framework that positions motion as a race against time and adversaries, engaging the reader in a high-speed chase toward clearly defined, winnable goals. It is characterised by entailments drawn from language about competitive races (i.e. ``accelerate,'' ``sprint,'' ``fast-track,'' ``outpace''). The race typically follows established rules and codes of conduct and requires rapid, strategic action that can be enhanced through training and preparation for success.

Its usage in AI policy may have several normative implications. Firstly, it may overemphasise speed and competition, leading to rushed decision-making and a lack of due diligence. This may be especially true to longer-term consequences described as `beyond the finish line', such as broader societal, economic or ethical consequences of `victory'. Secondly, its competitive nature may neglect collaborative or inclusive approaches that hinder comprehensive or inclusive development (i.e. hoarding knowledge). Thirdly, it may reinforce and legitimise existing power imbalances and lead to the concentration of decision-making power (and benefits) to a small number of actors with more resources, expertise and access who `pull away'. Lastly, focusing on a single race may lead to overlooking the importance of exploring alternative approaches, methodologies, or objectives in AI development.

Despite these risks, the \texttt{JOURNEY/RACE} frame may also serve to motivate and inspire action, driving innovation and progress in AI regulation. The structured nature of the competition could encourage adherence to established rules and codes of conduct, promoting responsible and ethical practices. The frame's emphasis on overcoming obstacles and achieving clearly defined goals may also provide a sense of purpose and direction in the face of complex challenges.

\Needspace{10\baselineskip}
\Needspace{10\baselineskip}
Examples include:
\begin{itemize}
\item ``\textbf{Accelerating towards} a cure in the \textbf{race against} the pandemic'',
\item ``In the \textbf{sprint} for market dominance, \textbf{speed} and innovation \textbf{lead}'',
\item ``We need to \textbf{overcome} \textbf{obstacles} to innovation'',
\item ``\textbf{Leading the pack} in adopting AI technologies'',
\item ``\textbf{Racing to the top} in green technology'',
\item ``The \textbf{fast track} to AI regulation compliance''.
\end{itemize}

As another common frame, \texttt{JOURNEY/RACE} shares the foundational aspect of movement and progression with \texttt{JOURNEY} but distinguishes itself through its emphasis on speed and competitive interaction, akin to \texttt{GAME}. It is often paired with \texttt{MACHINE} or \texttt{BODY} Narrative Frames, which facilitate conversations about the `vehicle' with which you are participating in the race. It contrasts with \texttt{WAR} by focusing more on individual or team achievement within a structured competition rather than conflict.

\section{Limitations}
This study is limited by its exclusive focus on English-language publications and the potential omission of relevant studies due to variations in terminology and indexing across databases. Future research should aim to expand the scope of the literature review and assess the inter-coder reliability of the proposed schema.

More broadly, as Quinn and Holland (1987) and Kövecses (2010) argue, capturing the full complexity of metaphorical framing is challenging due to cultural, contextual, and communicative factors that render metaphor interpretation inherently subjective. While acknowledging this debate, this paper does not seek to engage with it directly. Instead, it recognises the significant body of practice employing conceptual metaphor theory across various fields and argues that, given the widespread use of this approach and its potential to shape critical decisions in policy and AI development, it is valuable to attempt to bring these diverse practitioners together through a shared framework and common terms.

The proposed framework aims to facilitate the identification, critique, and situating of narrative frames in political rhetoric more easily while maintaining a strong connection to conceptual metaphor theory's foundations. By providing a common language and structure, this paper seeks to improve coordination and dialogue among the various fields of practice that employ conceptual metaphor theory, regardless of one's stance on the possibility of a truly objective typology. The goal is to enhance the coherence and impact of this work, given its growing influence on important societal issues.

\section{Conclusion}
This paper has identified significant concerns in current conceptual metaphor analysis practices, such as inconsistent definitions and conceptual understandings. Misalignment between different fields working on conceptual metaphor and AI further exacerbates these issues. To address these challenges, the Narrative Frames typology is proposed as a new approach to understanding and navigating the complex debate surrounding AI ethics and policy.

The Narrative Frames typology provides a shared language and framework for understanding the complex debate surrounding AI ethics and policy. It enables researchers, journalists, policymakers, and the public to more easily deconstruct the underlying assumptions, values, and goals promoted by different actors, leading to more informed discourse and decision-making. By providing a common foundation for interdisciplinary collaboration, Narrative Frames advance our understanding of how metaphors shape public perception and policy debates related to AI, which is crucial in a field where scholars from diverse backgrounds must work together.

The rigorous methodology, grounded in established theoretical foundations and practical applications, positions this paper as a meaningful contribution to the literature on metaphors, framing, and AI ethics. As the AI discourse evolves, the Narrative Frames categorisation system will serve as a valuable tool for navigating the complexities of this critical conversation.

\section*{References}
\begingroup
\small
\begin{sloppypar}
\setlength{\parindent}{-1.5em}
\setlength{\leftskip}{1.5em}
\setlength{\parskip}{0.3\baselineskip}

\noindent Agerri, R., Barnden, J., Lee, M., \& Wallington, A. (2008). Textual Entailment as an Evaluation Framework for Metaphor Resolution: A Proposal. In J. Bos \& R. Delmonte (Eds.), \textit{Semantics in Text Processing. STEP 2008 Conference Proceedings} (pp. 357--363). College Publications. \url{https://aclanthology.org/W08-2228}

\noindent Baker, C. F., Fillmore, C. J., \& Lowe, J. B. (1998). The Berkeley FrameNet Project. \textit{36th Annual Meeting of the Association for Computational Linguistics and 17th International Conference on Computational Linguistics, Volume 1}, 86--90. \url{https://doi.org/10.3115/980845.980860}

\noindent Barkhuizen, G., \& Wette, R. (2008). Narrative frames for investigating the experiences of language teachers. \textit{System}, \textit{36}(3), 372--387. \url{https://doi.org/10.1016/j.system.2008.02.002}

\noindent Borah, P. (2011). Conceptual Issues in Framing Theory: A Systematic Examination of a Decade’s Literature. \textit{Journal of Communication}, \textit{61}(2), 246--263. \url{https://doi.org/10.1111/j.1460-2466.2011.01539.x}

\noindent Bosman, J., \& Hagendoorn, L. (1991). Effects of Literal and Metaphorical Persuasive Messages. \textit{Metaphor and Symbolic Activity}, \textit{6}(4), 271--292. \url{https://doi.org/10.1207/s15327868ms0604_3}

\noindent Boulenger, V., Shtyrov, Y., \& Pulvermüller, F. (2012). When do you grasp the idea? MEG evidence for instantaneous idiom understanding. \textit{NeuroImage}, \textit{59}(4), 3502--3513. \url{https://doi.org/10.1016/j.neuroimage.2011.11.011}

\noindent Brugman, B. C., Burgers, C., \& Vis, B. (2019). Metaphorical framing in political discourse through words vs. concepts: A meta-analysis. \textit{Language and Cognition}, \textit{11}(1), 41--65. \url{https://doi.org/10.1017/langcog.2019.5}

\noindent Cacciatore, M. A., Scheufele, D. A., \& Iyengar, S. (2016). The End of Framing as we Know it \ldots{} and the Future of Media Effects. \textit{Mass Communication and Society}, \textit{19}(1), 7--23. \url{https://doi.org/10.1080/15205436.2015.1068811}

\noindent Cameron, L., Maslen, R., Todd, Z., Maule, J., Stratton, P., \& Stanley, N. (2009). The Discourse Dynamics Approach to Metaphor and Metaphor-Led Discourse Analysis. \textit{Metaphor and Symbol}, \textit{24}(2), 63.

\noindent Cammaerts, B. (2012). The strategic use of metaphors by political and media elites: The 2007--11 Belgian constitutional crisis. \textit{International Journal of Media and Cultural Politics}, \textit{8}(2/3).

\noindent Casasanto, D. (2009). Space for Thinking. In \textit{Language, Cognition and Space: State of the art and new directions} (pp. 453--478). Equinox Publishing. \url{https://www.equinoxpub.com/home/view-chapter/?id=22052}

\noindent Cave, S., \& Ó hÉigeartaigh, S. (2017). \textit{An AI Race for Strategic Advantage: Rhetoric and Risks} (SSRN Scholarly Paper 3446708). \url{https://papers.ssrn.com/abstract=3446708}

\noindent Chambers, N., \& Jurafsky, D. (2008). Unsupervised Learning of Narrative Event Chains. In J. D. Moore, S. Teufel, J. Allan, \& S. Furui (Eds.), \textit{Proceedings of ACL-08: HLT} (pp. 789--797). Association for Computational Linguistics. \url{https://aclanthology.org/P08-1090}

\noindent Charteris-Black, J. (2004). \textit{Corpus Approaches to Critical Metaphor Analysis}. Springer.

\noindent Charteris-Black, J. (2011). \textit{Politicians and Rhetoric: The Persuasive Power of Metaphor}. Springer.

\noindent Citron, F. M. M., \& Goldberg, A. E. (2014). Metaphorical sentences are more emotionally engaging than their literal counterparts. \textit{Journal of Cognitive Neuroscience}, \textit{26}(11), 2585--2595. \url{https://doi.org/10.1162/jocn_a_00654}

\noindent Colburn, T. R., \& Shute, G. M. (2008). Metaphor in computer science. \textit{Journal of Applied Logic}, \textit{6}(4), 526--533. \url{https://doi.org/10.1016/j.jal.2008.09.005}

\noindent Desai, R. H., Binder, J. R., Conant, L. L., Mano, Q. R., \& Seidenberg, M. S. (2011). The neural career of sensory-motor metaphors. \textit{Journal of Cognitive Neuroscience}, \textit{23}(9), 2376--2386. \url{https://doi.org/10.1162/jocn.2010.21596}

\noindent Entman, R. M. (1993). Framing: Toward Clarification of a Fractured Paradigm. \textit{Journal of Communication}, \textit{43}(4), 51--58. \url{https://doi.org/10.1111/j.1460-2466.1993.tb01304.x}

\noindent Entman, R. M., Matthes, J., \& Pellicano, L. (2008). Nature, Sources, and Effects of News Framing. In \textit{The Handbook of Journalism Studies}. Routledge.

\noindent Ezeifeka, C. (2013). \textit{Strategic Use of Metaphor in Nigerian Newspaper Reports: A Critical Perspective}. \url{https://www.semanticscholar.org/paper/Strategic-Use-of-Metaphor-in-Nigerian-Newspaper-A-Ezeifeka/b53c25d528d164bacbf37bf474d799e64a343b74}

\noindent Fauconnier, G., \& Turner, M. (2003). \textit{The Way We Think: Conceptual Blending And The Mind’s Hidden Complexities}. Basic Books.

\noindent Feldman, J., \& Narayanan, S. (2004). Embodied meaning in a neural theory of language. \textit{Brain and Language}, \textit{89}(2), 385--392. \url{https://doi.org/10.1016/S0093-934X(03)00355-9}

\noindent Feldman, R. (2006). From biological rhythms to social rhythms: Physiological precursors of mother-infant synchrony. \textit{Developmental Psychology}, \textit{42}(1), 175--188. \url{https://doi.org/10.1037/0012-1649.42.1.175}

\noindent Fillmore, C., \& Baker, C. F. (2001). \textit{Frame semantics for text understanding}. \url{https://www.semanticscholar.org/paper/Frame-semantics-for-text-understanding-Fillmore-Baker/72cc80486744320d5b9a2ea75e3fb1cde4ca669c}

\noindent Fillmore, C. J. (1975). An Alternative to Checklist Theories of Meaning. \textit{Annual Meeting of the Berkeley Linguistics Society}, 123--131. \url{https://doi.org/10.3765/bls.v1i0.2315}

\noindent Fillmore, C. J. (1976). Frame Semantics and the Nature of Language. \textit{Annals of the New York Academy of Sciences}, \textit{280}(1), 20--32. \url{https://doi.org/10.1111/j.1749-6632.1976.tb25467.x}

\noindent Fillmore, C. J., Johnson, C. R., \& Petruck, M. R. L. (2003). Background to FrameNet. \textit{International Journal of Lexicography}, \textit{16}(3), 235--250. \url{https://doi.org/10.1093/ijl/16.3.235}

\noindent Gibbs Jr., R. W. (2017). \textit{Metaphor wars: Conceptual metaphors in human life}. Cambridge University Press. \url{https://doi.org/10.1017/9781107762350}

\noindent Gibbs, R. W. (1994). \textit{The Poetics of Mind: Figurative Thought, Language, and Understanding}. Cambridge University Press.

\noindent Gibbs, R. W. (2011). Evaluating Conceptual Metaphor Theory. \textit{Discourse Processes}, \textit{48}(8), 529--562. \url{https://doi.org/10.1080/0163853X.2011.606103}

\noindent Goatly, A. (1997). \textit{The Language of Metaphors}. Routledge. \url{https://doi.org/10.4324/9780203210000}

\noindent Hamrita, M. (2016). \textit{The Metaphorical and Ideological Representation of the Political Opponent in the Hardline Islamist Discourse in Tunisia}. \url{https://www.academia.edu/66645220/The_Metaphorical_and_Ideological_Representation_of_the_Political_Opponent_in_the_Hardline_Islamist_Discourse_in_Tunisia}

\noindent Hanna, R. (2017). Kant’s Theory of Judgment: Completing the Picture of Kant’s Metaphysics of Judgment. In \textit{Stanford Encyclopedia of Philosophy}. \url{https://plato.stanford.edu/entries/kant-judgment/supplement5.html}

\noindent Herrnson, P. S., Stokes-Brown, A. K., \& Hindman, M. (2007). Campaign Politics and the Digital Divide: Constituency Characteristics, Strategic Considerations, and Candidate Internet Use in State Legislative Elections. \textit{Political Research Quarterly}, \textit{60}(1), 31--42. \url{https://doi.org/10.1177/1065912906298527}

\noindent Hertog, J., \& Mcleod, D. (2001). A multiperspectival approach to framing analysis: A field guide. In \textit{Reese, Gandy, and Grant} (pp. 139--161).

\noindent Holland, D., \& Quinn, N. (Eds.). (1987). \textit{Cultural Models in Language and Thought}. Cambridge University Press.

\noindent Hsieh, H.-F., \& Shannon, S. E. (2005). Three Approaches to Qualitative Content Analysis. \textit{Qualitative Health Research}, \textit{15}(9), 1277--1288. \url{https://doi.org/10.1177/1049732305276687}

\noindent Imani, A. (2022). Critical Metaphor Analysis: A Systematic Step-by-step Guideline. \textit{LSP International Journal}, \textit{9}(1). \url{https://doi.org/10.11113/lspi.v9.17975}

\noindent Jiang, J., Lupoiu, R., Wang, E. W., Sell, D., Hugonin, J. P., Lalanne, P., \& Fan, J. A. (2020). MetaNet: A new paradigm for data sharing in photonics research. \textit{Optics Express}, \textit{28}(9), 13670--13681. \url{https://doi.org/10.1364/OE.388378}

\noindent Johnson, M. (1987). \textit{The body in the mind: The bodily basis of meaning, imagination, and reason}. University of Chicago Press.

\noindent Kant, I. (2003). \textit{The Critique of Pure Reason} [eBook]. Project Gutenberg. \url{https://www.gutenberg.org/files/4280/4280-h/4280-h.htm}

\noindent Kovecses, Z. (2010). \textit{Metaphor: A Practical Introduction}. Oxford University Press.

\noindent Lacey, S., Stilla, R., \& Sathian, K. (2012). Metaphorically feeling: Comprehending textural metaphors activates somatosensory cortex. \textit{Brain and Language}, \textit{120}(3), 416--421. \url{https://doi.org/10.1016/j.bandl.2011.12.016}

\noindent Lakoff, G. (2004). \textit{Don’t Think of an Elephant: Know Your Values and Frame the Debate}. Chelsea Green Publishing.

\noindent Lakoff, G. (2008). The neural theory of metaphor. In \textit{The Cambridge handbook of metaphor and thought} (pp. 17--38). Cambridge University Press. \url{https://doi.org/10.1017/CBO9780511816802.003}

\noindent Lakoff, G. (2009). \textit{The Political Mind: A Cognitive Scientist’s Guide to Your Brain and Its Politics}. Penguin Books.

\noindent Lakoff, G., \& Johnson, M. (1980). \textit{Metaphors We Live By}. University of Chicago Press.

\noindent Landau, M. J., Oyserman, D., Keefer, L. A., \& Smith, G. C. (2014). The college journey and academic engagement: How metaphor use enhances identity-based motivation. \textit{Journal of Personality and Social Psychology}, \textit{106}(5), 679--698. \url{https://doi.org/10.1037/a0036414}

\noindent Lee, M., \& Barnden, J. (2001a). \textit{Application of the ATT-Meta Metaphor-Understanding System to an Example of the Metaphorical View of MIND PARTS AS PERSONS}.

\noindent Lee, M., \& Barnden, J. (2001b). \textit{Mental Metaphors from the Master Metaphor List: Empirical Examples and the Application of the ATT-Meta System}.

\noindent Leeper, T., \& Slothuus, R. (2020). How the News Media Persuades: Framing Effects and Beyond. In E. Suhay, B. Grofman, \& A. H. Trechsel (Eds.), \textit{Oxford Handbook of Electoral Persuasion} (pp. 151--168). Oxford University Press. \url{https://doi.org/10.1093/oxfordhb/9780190860806.013.4}

\noindent Litkowski, K. (2010). Multilingual FrameNets in Computational Lexicography: Methods and Applications. \textit{International Journal of Lexicography}, \textit{23}, 105--109. \url{https://doi.org/10.1093/ijl/ecp034}

\noindent Martin, J. H. (1994). Metabank: A Knowledge-Base of Metaphoric Language Conventions. \textit{Computational Intelligence}, \textit{10}(2), 134--149. \url{https://doi.org/10.1111/j.1467-8640.1994.tb00161.x}

\noindent Matthes, J. (2009). What’s in a Frame? A Content Analysis of Media Framing Studies in the World’s Leading Communication Journals, 1990--2005. \textit{Journalism \& Mass Communication Quarterly}, \textit{86}(2), 349--367. \url{https://doi.org/10.1177/107769900908600206}

\noindent Minsky, M. (1975). A Framework for Representing Knowledge. In P. Winston (Ed.), \textit{The Psychology of Computer Vision} (pp. 211--277). McGraw-Hill. \url{https://dspace.mit.edu/bitstream/handle/1721.1/6089/AIM-306.pdf?sequence=2}

\noindent Musolff, A. (2004). \textit{Metaphor and political discourse: Analogical reasoning in debates about Europe}. \url{https://doi.org/10.1057/9780230504516}

\noindent Musolff, A. (2006). Metaphor Scenarios in Public Discourse. \textit{Metaphor and Symbol}, \textit{21}. \url{https://doi.org/10.1207/s15327868ms2101_2}

\noindent Musolff, A. (2011). Metaphor in political dialogue. \textit{Language and Dialogue}, \textit{1}, 191--206. \url{https://doi.org/10.1075/ld.1.2.02mus}

\noindent Needham-Didsbury, I. (2013). Endangered Metaphors. \textit{Intercultural Pragmatics}. \url{https://www.academia.edu/5236298/Idström_and_Piirainen_2012_Endangered_Metaphors_Cognitive_Linguistic_Studies_in_Cultural_Contexts_CLSCC_}

\noindent Nikolí, D. (2009). \textit{Is synaesthesia actually ideaesthesia? An inquiry into the nature of the phenomenon}. \url{https://www.semanticscholar.org/paper/IS-SYNAESTHESIA-ACTUALLY-IDEAESTESIA-AN-INQUIRY-THE-Nikolí/622600e08f7d1ee2c6e27a8933239addbc56bd44}

\noindent Piaget, J. (1952). \textit{The Origins of Intelligence in Children}. International Universities Press.

\noindent Pragglejaz Group. (2007). MIP: A Method for Identifying Metaphorically Used Words in Discourse. \textit{Metaphor and Symbol}, \textit{22}(1), 1--39. \url{https://doi.org/10.1080/10926480709336752}

\noindent Robins, S., \& Mayer, R. E. (2000). The Metaphor Framing Effect: Metaphorical Reasoning About Text-Based Dilemmas. \textit{Discourse Processes}, \textit{30}(1), 57--86. \url{https://doi.org/10.1207/S15326950dp3001_03}

\noindent Schank, R. C., \& Abelson, R. P. (1977). \textit{Scripts, Plans, Goals, and Understanding: An Inquiry Into Human Knowledge Structures}. Psychology Press. \url{https://doi.org/10.4324/9780203781036}

\noindent Scheufele, D., \& Iyengar, S. (2014). \textit{The state of framing research: A call for new directions}. \url{https://doi.org/10.1093/oxfordhb/9780199793471.013.47}

\noindent Schmidt, G. L., \& Seger, C. A. (2009). Neural correlates of metaphor processing: The roles of figurativeness, familiarity and difficulty. \textit{Brain and Cognition}, \textit{71}(3), 375--386. \url{https://doi.org/10.1016/j.bandc.2009.06.001}

\noindent Semino, E. (2008). \textit{Metaphor in Discourse}. Cambridge University Press.

\noindent Semino, E., Demjén, Z., \& Demmen, J. (2018). An Integrated Approach to Metaphor and Framing in Cognition, Discourse, and Practice, with an Application to Metaphors for Cancer. \textit{Applied Linguistics}, \textit{39}(5), 625--645. \url{https://doi.org/10.1093/applin/amw028}

\noindent Shutova, E., Kiela, D., \& Maillard, J. (2016). Black Holes and White Rabbits: Metaphor Identification with Visual Features. In K. Knight, A. Nenkova, \& O. Rambow (Eds.), \textit{Proceedings of the 2016 Conference of the North American Chapter of the Association for Computational Linguistics: Human Language Technologies} (pp. 160--170). \url{https://doi.org/10.18653/v1/N16-1020}

\noindent Steen, G. (2011). The Language of Knowledge Management: A Linguistic Approach to Metaphor Analysis. \textit{Systems Research and Behavioral Science}, \textit{28}, 181--188. \url{https://doi.org/10.1002/sres.1087}

\noindent Steinhart, E. (2005). Generating and Interpreting Metaphors with Netmet. \textit{APA Newsletter on Philosophy and Computers}, \textit{4}(2). \url{https://philarchive.org/rec/STEGAI-2}

\noindent Stickles, E., David, O., Dodge, E., \& Hong, J. (2016). Formalizing contemporary conceptual metaphor theory: A structured repository for metaphor analysis. \textit{Constructions and Frames}, \textit{8}. \url{https://doi.org/10.1075/cf.8.2.03sti}

\noindent Taiwo, R. (2013). Metaphor in Nigerian Political Discourse. \textit{Select Papers from the 2008 Stockholm Metaphor Festival}, 193--207. \url{https://www.academia.edu/14941242/Metaphor_in_Nigerian_Political_Discourse}

\noindent Thibodeau, P. H., \& Boroditsky, L. (2013). Natural Language Metaphors Covertly Influence Reasoning. \textit{PLOS ONE}, \textit{8}(1), e52961. \url{https://doi.org/10.1371/journal.pone.0052961}

\noindent Thibodeau, P. H., Hendricks, R. K., \& Boroditsky, L. (2017). How Linguistic Metaphor Scaffolds Reasoning. \textit{Trends in Cognitive Sciences}, \textit{21}(11), 852--863. \url{https://doi.org/10.1016/j.tics.2017.07.001}

\noindent Thibodeau, P., Winneg, A., Frantz, C., \& Flusberg, S. (2016). The mind is an ecosystem: Systemic metaphors promote systems thinking. \textit{Metaphor and the Social World}, \textit{6}, 225--242. \url{https://doi.org/10.1075/msw.6.2.03thi}

\noindent Veale, T., Shutova, E., \& Klebanov, B. (2016). Metaphor: A Computational Perspective. \textit{Synthesis Lectures on Human Language Technologies}, \textit{9}, 1--160. \url{https://doi.org/10.2200/S00694ED1V01Y201601HLT031}

\noindent Vervaeke, J., \& Kennedy, J. M. (1996). Metaphors in Language and Thought: Falsification and Multiple Meanings. \textit{Metaphor and Symbolic Activity}, \textit{11}(4), 273--284. \url{https://doi.org/10.1207/s15327868ms1104_3}

\noindent Ziem, A. (2021). \textit{FrameNet, Barsalou Frames and the Case of Associative Anaphora} (pp. 93--112). \url{https://doi.org/10.1515/9783110720129-005}

\end{sloppypar}
\endgroup

\end{document}